OPEN

# Dielectric characterisation of human adrenal glands and adrenal tumours for the development of microwave ablation technologies for hypertension treatment

Bilal Amin[1,2,3]✉, Adnan Elahi[1,2], Grazia Cappiello[1,3], Niko Ištuk[1,2], Eoghan Dunne[1,2,3], Finn Krewer[1,2], Anne Marie Quinn[4], Aoife Lowery[5], Punit Prakash[6], Michael Conall Dennedy[3] & Martin O'Halloran[1,2,3]

Adrenal gland-induced hypertension results from underlying adrenal gland disorders including Conn's syndrome, Cushing's syndrome, and Pheochromocytoma. These adrenal disorders are a risk for cardiovascular and renal morbidity and mortality. Clinically, treatment for adrenal gland-induced hypertension involves a pharmaceutical or surgical approach. The former presents very significant side effects whereas the latter can be ineffective in cases where the adrenal disorder reoccurs in the remaining contralateral adrenal gland. Due to the limitations of existing treatment methods, minimally invasive treatment options like microwave ablation (MWA) have received significant attention for treating adrenal gland disorders. A precise comprehension of the dielectric properties of human adrenal glands will help to tailor energy delivery for MWA therapy, thus offering the potential to optimise treatments and minimise damage to surrounding tissues. This study reports the ex vivo dielectric properties of human adrenal glands, including the cortex, medulla, capsule, and tumours, based on the data obtained from four patients (diagnosed with Conn's syndrome, Cushing's syndrome, and Pheochromocytoma) who underwent unilateral adrenalectomy at the University Hospital Galway, Ireland. An open-ended coaxial probe measurement technique was used to measure the dielectric properties for a frequency range of 0.5–8.5 GHz. The dielectric properties were fitted using a two-pole Debye model, and a weighted least squares method was employed to optimise the model parameters. Moreover, the dielectric properties of adrenal tissues and tumours were compared across frequencies commonly used in MWA, including 915 MHz, 2.45 GHz, and 5.8 GHz. The study found that the dielectric properties of adrenal tumours were influenced by the presence of lipid-rich adenomas, and the dielectric properties of Cushing's syndrome tumour were lowest in comparison to the tumours in patients diagnosed with Conn's syndrome and Pheochromocytoma. Furthermore, a notable difference was observed in the dielectric properties of the medulla and cortex among patients diagnosed with Conn's syndrome, Cushing's syndrome, and Pheochromocytoma. These findings have significant implications for the diagnosis and treatment of adrenal tumours, including the optimisation of MWA therapy for precise ablation of adrenal masses.

[1]Translational Medical Device Laboratory, Lambe Institute for Translational Research & HRB Clinical Research Facility, University Hospital Galway, Galway H91 YR71, Ireland. [2]Electrical and Electronic Engineering, University of Galway, Galway H91 TK33, Ireland. [3]School of Medicine, University of Galway, Galway H91 TK33, Ireland. [4]Department of Anatomic Pathology, University Hospital Galway, Galway H91 TK33, Ireland. [5]Discipline of Surgery, Lambe Institute for Translational Research, University of Galway, Galway H91 TK33, Ireland. [6]Department of Electrical and Computer Engineering, Kansas State University, Manhattan, KS 66506, USA. ✉email: bilal.amin@universityofgalway.ie

Hypertension is a highly prevalent health condition worldwide[1]. According to the World Health Organization, in 2021, approximately 1.28 billion people aged 30 to 79 years have hypertension globally, with a prevalence rate of around 22%[2,3]. Moreover, hypertension is responsible for approximately 10 million deaths every year globally, and it is a significant risk factor for cardiovascular diseases, stroke, and kidney failure[2]. The economic burden of hypertension is also significant, with estimates suggesting that it costs around $370 billion globally each year in healthcare expenses and lost productivity[4]. This burden is expected to increase in line with both the ageing population and the rising costs of diagnostic and therapeutic technologies[4].

Adrenal gland-induced hypertension, also known as secondary hypertension, is one of the leading curable causes of hypertension resulting from abnormalities in the adrenal glands[5]. Adrenal endocrinopathies collectively represent the most common identifiable secondary cause, often caused by primary aldosteronism (PA), and rarely Cushing syndrome and Pheochromocytoma[5]. Within this cluster of diseases, PA and Cushing disease arise from the adrenal cortex, whereas phaeochromocytoma arises from the inner medulla[6]. PA accounts for 6–12% of all hypertension, and as high as 22% of so-called resistant hypertension[5]. Cushing syndrome is a multimorbid systemic disease, which carries an increased risk of cardiovascular disease, cancer, osteoporosis, and infection[7]. Phaeochromocytoma is typically associated with severe treatment-resistant hypertension and carries a higher cardiovascular risk when compared to primary hypertension[8]. PA is typically characterised by hypertension which is resistant to standard pharmacotherapy and is also associated with a higher risk of atrial fibrillation, stroke, and myocardial infarction compared with blood pressure-matched hypertension[5]. Adrenal lesions underlying Cushing Syndrome and Phaeochromocytoma are typically unilateral, large, and easily identifiable and are treated surgically with unilateral adrenalectomy[9]. There are selected case studies, which have also used radiofrequency (RFA) to treat both Cushing's and Phaeochromocytoma[5]. Although in patients with Phaeochromocytoma, this approach carries the risk of hypertensive crisis and consequently should not be routinely undertaken, should never be undertaken in large lesions, and must be done under appropriate anti-adrenergic pre-treatment.

PA is caused by unilateral aldosterone-producing adenomas (APA) in 30–40% of all cases. In the remaining cases, it is caused bilaterally by aldosterone-producing nodules, micronodules, or hyperplasia. In this regard, APAs are typically treated with unilateral adrenalectomy and bilateral disease is treated with anti-hypertensive pharmacotherapy, primarily with mineralocorticoid receptor antagonists (MRA)[5]. Bilateral adrenalectomy is rarely undertaken for bilateral causes of PA because of the unacceptable post-procedural complication of adrenocortical insufficiency. Hence more than 60% of all patients with PA remain on life-long pharmacotherapy with no option for definitive management of this endocrinopathy[10]. Advances in the localisation of aldosterone-producing adrenocortical lesions using molecular imaging offer the potential of directing more selective intervention to definitively manage both unilateral and bilateral PA. These interventions include selective adrenalectomy and minimally invasive techniques such as thermal ablation[10]. Where bilateral intervention is undertaken, the pathologic tissue must be appropriately ablated or removed. However, it is equally important that the normally functioning adrenocortical tissue is preserved to avoid post-procedural adrenocortical insufficiency.

Microwave ablation (MWA) offers the potential for treating APA and aldosterone-producing nodules with a high degree of accuracy and precision. The pattern of ablation exhibited by MWA applicator probes offers more uniform and controlled heating of the target tissue and directionality when compared to RFA, thus offering a more precise ablation zone confined to the diseased part of the gland/nodule, leading to better preservation of surrounding healthy tissues[11–13]. Moreover, MWA has a higher thermal efficiency and can achieve higher temperatures more rapidly compared to RFA or cryoablation[12]. This results in shorter procedure times and more effective tissue destruction. Additionally, MWA is less susceptible to the "heat sink" effect caused by nearby blood vessels, making it more effective in highly vascularised tissues[13]. In terms of safety, MWA has been associated with lower complication rates and less post-procedural pain compared to other ablation techniques. Limited clinical case series of MWA have shown promising results as a minimally invasive treatment option for adrenal tumours and other adrenal diseases, offering several benefits over traditional adrenalectomy or RFA, such as precise targeting, efficient ablation, and reduced risk of complications[13,14]. In a MWA procedure, the ablation antenna, usually enclosed in a needle or stiff catheter, is inserted into the target area under the guidance of ultrasound or computed tomography (CT) imaging[15,16]. The microwave energy emitted by the antenna raises the temperature of surrounding tissues to 50–60 °C or above, which leads to cell death by the mechanism of coagulation necrosis[14]. Numerous clinical studies have documented the use of MWA to treat tumours in the liver, lung, kidney, and breast[17]. Currently, available MWA systems in clinical use operate at frequencies of 915 MHz, 2.45 GHz, or 5.8 GHz[18,19]. The latter frequency provides faster heating and shorter antenna lengths. In previous studies, the authors have demonstrated that the use of selective MWA within the adrenal gland can be used to target areas of 1 $cm^3$ or below (the typical size of an APA or aldosterone-producing nodule), in both adrenal glands, while preserving the adjacent adrenal cortex and without inducing post-procedural adrenocortical insufficiency[20]. It has been demonstrated that sublethal hyperthermia at temperatures expected within the transitional zone between ablated and non-ablated tissue induces only transient inhibition of steroidogenesis[20]. The development of precise MWA devices which selectively target pathological tissue types requires a comprehensive knowledge of the dielectric properties of both normal and pathological tissue.

Dielectric properties play an important role in determining the efficiency and effectiveness of the MWA treatment of all tissues including adenomas of the adrenal gland[15,18,21]. Overall, dielectric tissue properties determine how they interact with the microwave energy emitted by a thermal ablation antenna, affecting the amount of energy absorbed and the resulting heating[22]. Adrenal tumours typically have low to intermediate dielectric values, which can result in more efficient absorption of microwave energy and better compared to heating[13,15,18].

However, the surrounding tissues, such as blood vessels and muscle, have higher dielectric values and can act as barriers to microwave energy transmission, reducing the amount of energy that reaches the tumour[13]. This reduction in delivered energy can result in a less- effective ablation and in the need for higher energy levels and longer treatment times to be used. Therefore, understanding the dielectric properties of adrenal tumours and surrounding tissues is essential for the effective planning and execution of MWA procedures. Accurate characterisation of these properties can therefore inform optimal energy levels and antenna placement for efficient and effective ablation, to improve treatment outcomes for patients[15,23,24].

Significant experimental work has been performed on characterising the dielectric properties of various biological tissues including breast, tumour, lung, kidney, skin, bone, and heart among others[24–27]. Although a large number of experimental studies were conducted acquiring the dielectric properties of various biological tissues, to our knowledge only three studies have reported the dielectric properties of adrenal glands with only one study reporting properties of human adrenal glands[18,21,28]. Peyman et al.[28] reported the in vivo dielectric properties of the outer layer of the porcine adrenal gland ($N=12$), referred to as the adrenal cortex, for a frequency range of 50 MHz–20 GHz. However, it is important to note that this study did not report the dielectric properties of the adrenal medulla, which limits its applicability for developing comprehensive anatomically and electrically accurate computer models of adrenal glands. Shahzad et al.[21] reported the ex vivo dielectric properties of bovine adrenal glands ($N=12$), for a frequency range of 500 MHz–20 GHz. This study reported properties of both the adrenal cortex and the medulla. However, it is important to note that this work was conducted on animal tissues, which may limit the direct applicability of the findings to human tissues. The differences in tissue composition and structure between bovine and human adrenal glands could influence the dielectric properties, and thus, further studies on human tissues are necessary to validate and extend these results. Bottiglieri et al.[18] reported the ex vivo dielectric properties of ovine ($N=8$) and human ($N=6$) adrenal glands for a frequency range of 500 MHz–8.5 GHz. In this study, the authors were unable to examine the inner portions of the human adrenal glands, specifically the medulla, and their findings only pertain to the normal adrenal cortex of human glands. Moreover, the ex vivo studies may not perfectly represent in vivo conditions particularly when the time from excision to measurement is long. Further, the literature suggested that there is a noticeable lack of dielectric properties data on adrenal tumours. Therefore, it is also necessary to characterise the dielectric properties of adrenal tumours to develop and optimise MWA-based selective treatment of shallow APAs and aldosterone-secreting nodules.

This study presents new data which outlines the detailed human dielectric properties of the normal adrenal gland which includes the capsule, cortex, and medulla. We also describe the dielectric properties of APA and compare these with non-aldosterone-producing adrenocortical adenomas (NAPACA) causing Cushing Syndrome and Phaeochromocytoma. The data presented herein will help inform treatment planning and future applicator probe design for improved selective, precision ablation of APA. To this end, a total of four adrenal gland samples were acquired from four different patients who have undergone adrenalectomy at the University Hospital Galway, Ireland. Two of the patients included in the study were diagnosed with Cushing's syndrome, while the other two were diagnosed with PA and Pheochromocytoma. The samples were sectioned by the pathologist to access the medulla, cortex, and capsule. The pathologist highlighted the tumour regions within each adrenal gland. Once the adrenal gland samples were sectioned, the ex vivo dielectric properties were measured using an open-ended coaxial probe across the microwave frequency range of 0.5–8.5 GHz. To the best of the knowledge of the authors, this study is the first to document the dielectric properties of the human adrenal medulla, capsule, and adrenal tumour from patients diagnosed with Cushing syndrome in the literature. Moreover, the dielectric properties of tumours from patients diagnosed with Conn's syndrome and Pheochromocytoma measured in this study are compared with the dielectric properties of human adrenal glands reported by Bottiglieri et al.[18]. The knowledge of dielectric properties acquired from patients diagnosed with different adrenal disorders will help in the development and optimisation of MWA-based technologies for hypertension treatment. By understanding the dielectric properties of tissues affected by hypertension, medical professionals can better predict how those tissues will respond to electromagnetic (EM) energy used in MWA-based treatments. This knowledge can help to optimise treatment parameters and improve treatment efficacy and safety[18]. Furthermore, comparing the dielectric properties of healthy and diseased tissues could facilitate the development of more targeted and personalised MWA-based treatments for hypertension.

The remainder of this paper is structured as follows: section "Materials and methods" discusses the source of human adrenal gland samples, the methodology used to measure the dielectric properties of adrenal glands, the measurement protocol and system uncertainty analysis of the dielectric properties measurement equipment and the fitting model used on dielectric properties data. Section "Results and discussion" presents the results and discussion on the dielectric properties measurement of ex vivo human adrenal glands including the medulla, cortex, capsule, and tumours. Finally, conclusions are drawn in section "Conclusions".

## Materials and methods
### Source of tissues
The human adrenal glands were acquired from a total of four patients ($N=4$) with informed consent. The ethical approval for the experiments was obtained from the Clinical Research Ethics Committee, Galway University Hospitals, Galway, Ireland. The patients were diagnosed with adrenal disorders including Conn's syndrome ($N=1$), Cushing's syndrome ($N=2$), and Pheochromocytoma ($N=1$). The mean age of patients was $48.75 \pm 14.14$ years. Two of the four patients were female. The unilateral adrenalectomy of each patient was performed at the University Hospital Galway, Ireland using a retroperitoneoscopic approach. During the unilateral adrenalectomy of each of the four patients, the entire adrenal gland was excised including the adrenal cortex, medulla, and tumour. All methods were performed in accordance with the relevant guidelines and regulations. Two of the patients had left unilateral retroperitoneoscopic adrenalectomy whereas the other two patients had right unilateral adrenalectomy. The pre-operative radiological dimensions of the tumour varied among the patients with different diagnoses. For the patient diagnosed with Conn's syndrome, the tumour size measured 9 × 7 × 4 mm. Meanwhile, the two patients diagnosed with Cushing's syndrome presented with tumour sizes of 55 × 45 × 15 mm and 34 × 15 × 21 mm, respectively. As for the patient diagnosed with Pheochromocytoma, the tumour size measured 52 × 41 × 46 mm. After each unilateral adrenalectomy in the surgical room, the samples were enclosed in a closed plastic container and transported to the pathology room of the University Hospital Galway, Ireland. The approximate time elapsed between the surgery and the dielectric properties measurement was less than 10 min to ensure ex vivo measurements closely represent in vivo measurements. The samples were not preserved in any preservative liquid during the time between surgery and dielectric properties measurements. The microscopic pathology features associated with the adrenal gland samples obtained from each of the patients are shown in Fig. 1.

## Dielectric properties measurements

The dielectric properties were measured using a slim-form open-ended coaxial probe (Keysight 85070E, Santa Rosa, CA, US). A vector network analyser (Keysight VNA E5063A, Santa Rosa, CA, US) was used to record the reflection coefficient ($S_{11}$) at the calibration plane of the probe. Data were recorded in the range of 0.5–8.5 GHz over 101 linearly spaced frequency points. A software tool (Keysight N1500A, Santa Rosa, CA, US) was used to convert the $S_{11}$ parameters to real ($\varepsilon'(\omega)$) and imaginary ($\varepsilon''(\omega)$) parts of the complex permittivity[18,24,29].

The dielectric properties were measured as shown in Fig. 2. The slim-form probe was firmly attached to the VNA to minimise the measurement inaccuracies caused by repositioning the probe and to ensure that the probe remained stable while in contact with the sample and that the movement of the probe did not cause any artefacts during the dielectric properties measurements. Sample temperature during measurement was recorded as between 18 and 20 °C with a digital infrared thermometer (N85FR).

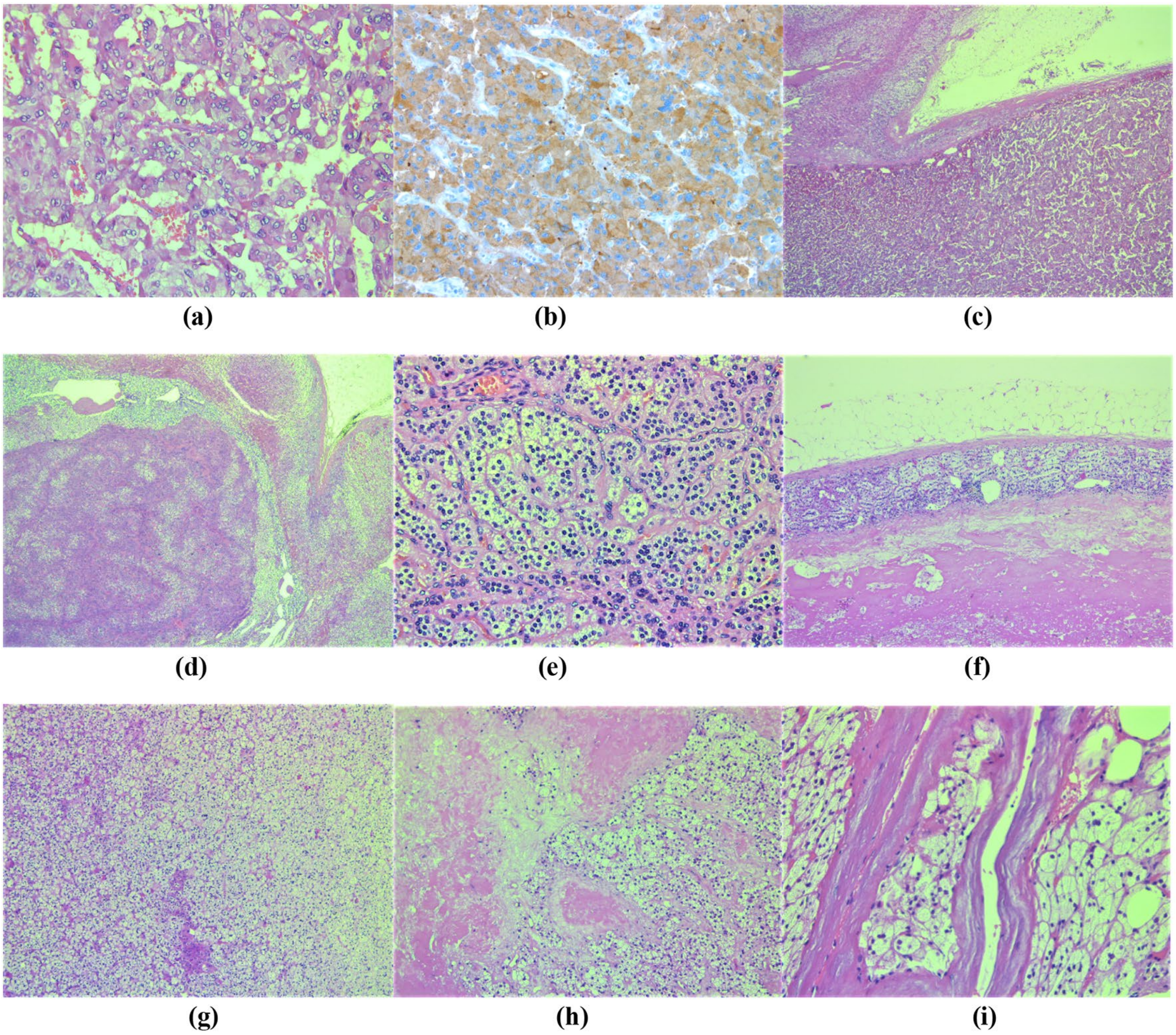

**Fig. 1.** (**a**) Phaeochromocytoma cells, 200 × H&E (**b**) Phaeochromocytoma cells, 200 × synaptophysin (**c**) Phaeochromocytoma bottom with normal adrenal gland above, 25 × H&E (**d**) adrenal cortical adenoma (clinically Conn's syndrome), 25 × H&E (**e**) same adenoma as d at 200 × magnification with clear fasciculate-type cells and some eosinophilic glomerulosa-type cells (**f**) different necrotic cortical adenoma (clinically Cushing's syndrome) with compressed peripheral rim of normal tissue, 50 × H&E (**g**) 3rd adrenal cortical adenoma (clinically Cushing's) with areas of diffuse clear cell growth, 50 × H&E (**h**) same adenoma with foci of confluent necrosis, 100 × H&E (**i**) same adenoma with possible venous invasion 200 × H&E.

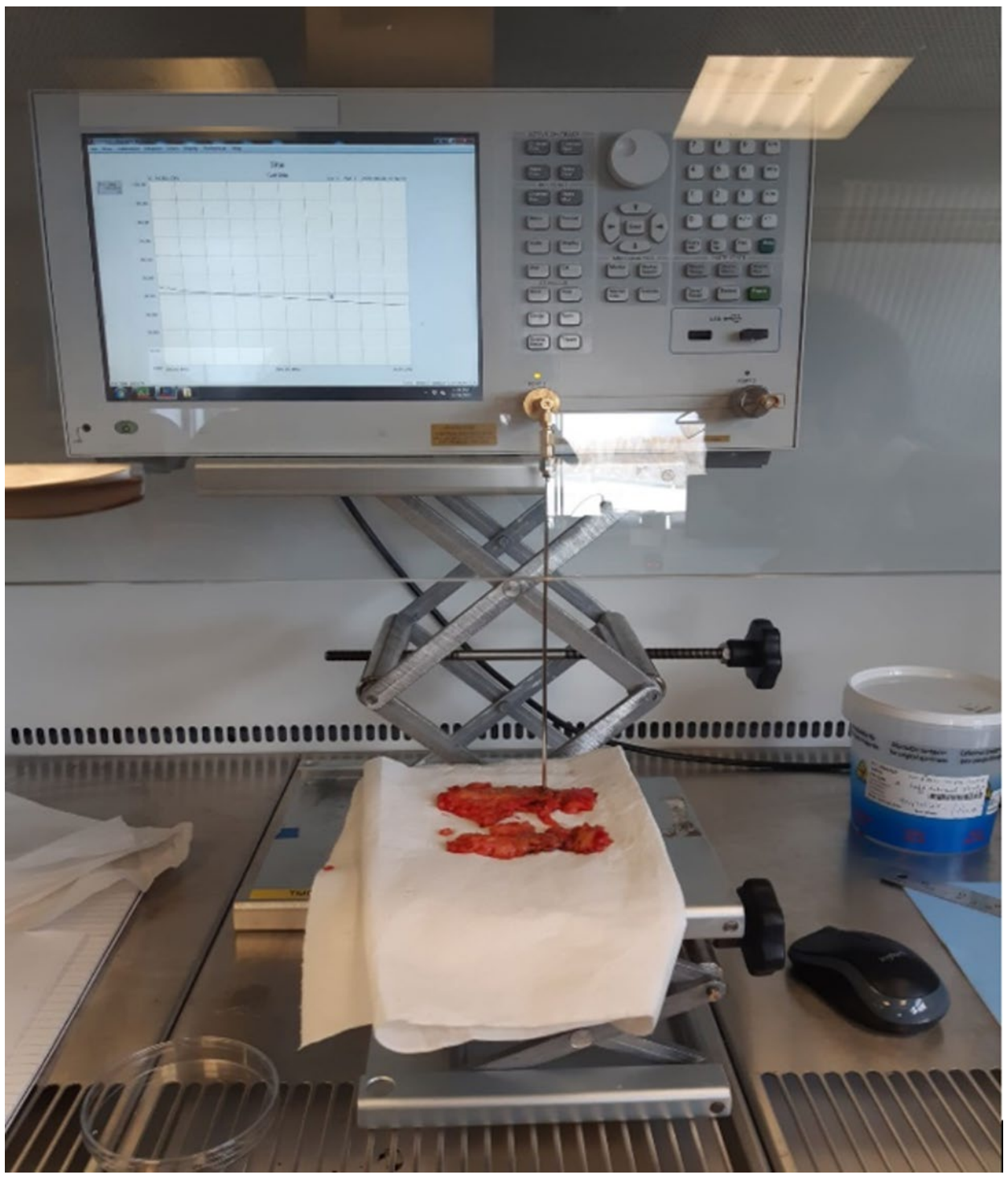

**Fig. 2.** Dielectric properties measurement setup.

The ex vivo dielectric properties of human adrenal glands were measured in the pathology laboratory of the hospital. The measurement location of the medulla, cortex, capsule, and tumour was univocally identified by the pathologist before taking the dielectric properties measurements. At each measurement location of the sample, a total of three measurements were performed. Additionally, histological confirmation was performed to ensure the accuracy of the tissue identification and to validate the integrity of the measured properties. Accordingly, for the four patients, a total of 192 dielectric measurements were recorded and analysed. The probe was cleaned with an alcohol wipe before each measurement. The probe was placed at maximum locations of a given tissue type, depending on the size of the sample. The number of measurement locations for each of the diagnoses with corresponding tissue types is listed in Table 1.

## Measurement protocol and system uncertainty analysis

Measurement equipment was calibrated before each measurement session. Calibration of the equipment was performed with three standard loads (open (air), short (shorting block), and load (deionized water))[29,30]. The temperature of the deionized water during each measurement session was 19.55 ± 1 °C. The calibration was verified with 6 validation measurements of the dielectric properties of 0.1 M NaCl (saline) solution at 21.6 °C[29,31]. An estimation of the uncertainty of the measurement accuracy was performed for each frequency point and averaged over the frequency range. The uncertainty of the accuracy in terms of percentage is defined as:

$$UC_{ACC}(f) = \left( \frac{DP_{meas}(f) - DP_{ref}(f)}{DP_{ref}(f)} \right) \times 100 \quad (1)$$

| Diagnosis | No. of samples (*N*) | Number of dielectric properties measurement points | | | | |
|---|---|---|---|---|---|---|
| | | Medulla | Cortex | Capsule | Tumour | Total |
| Conn's syndrome | 1 | 7 | 5 | 3 | 3 | 54 |
| Cushing's syndrome | 2 | 12 | 6 | – | 20 | 114 |
| Pheochromocytoma | 1 | 3 | 2 | – | 3 | 24 |

**Table 1.** Diagnosis and number of dielectric properties measurement points for each tissue type.

where $DP_{meas}(f)$ represents the measured dielectric properties of 0.1 M NaCl solution at a specific frequency point $f$ and $DP_{ref}(f)$ represents the reference dielectric properties of 0.1 M NaCl solution at the measured temperature[32]. The uncertainty related to repeatability ($UC_{REP}$) provides a measure of the measurement error. The uncertainty in repeatability of measurements expressed as a percentage is:

$$UC_{REP}(f) = \left( \frac{DP_{meas}(f) - DP_{mean}(f)}{DP_{mean}(f)} \right) \times 100 \tag{2}$$

where $DP_{mean}(f)$ represents the mean of the measured dielectric properties of 0.1 M NaCl solution at a specific frequency point $f$. Uncertainty from cable movement was not considered as the probe was directly connected to the VNA. Combined uncertainty ($UC_{COM}$) was calculated using the standard uncertainty ($UC_{STD}$) of the uncertainty components ($UC_{ACC}$ and $UC_{REP}$). The expanded uncertainty ($UC_{EXP}$) was calculated as a 95% confidence interval, in line with the guidelines of the National Institute of Standard and Technology (NIST)[33]. The expanded uncertainty was found to be 1.04% and 5.59% for relative permittivity and conductivity respectively (Table 2). The uncertainty components are computed separately for both the relative permittivity and conductivity of 0.1 M NaCl for a frequency range of 0.5–8.5 GHz.

## Fitting model on ex vivo human dielectric properties data

Dielectric properties of biological tissues are frequency-dependent[34]. The Cole–Cole model and the Debye model[25,35] are two commonly used parametric models of dielectric properties. The Cole–Cole model is computationally more complex than the Debye model and cannot be easily expressed in the time domain for tissue dielectric simulations[25]. The Debye model can be easily expressed in both time and frequency domains and is less computationally expensive[25]. The Debye model parameters are widely used in EM numerical simulations[33]. Hence, the Debye model is used for fitting and representing the data in this study. The Debye model is defined as:

$$\hat{\varepsilon} = \varepsilon_\infty + \sum_{p=1}^{n} \frac{\Delta\varepsilon_p}{1 + j\omega\tau_p} + \frac{\sigma_s}{j\omega\varepsilon_o}, \tag{3}$$

where $\hat{\varepsilon}$ is the complex permittivity, $\varepsilon_\infty$ is the permittivity at the highest frequency, $\Delta\varepsilon_p$ is the change in the permittivity at the $p^{th}$ dispersion, $\tau_p$ is the relaxation constant at the $p^{th}$ dispersion, $\sigma_s$ is the static ionic conductivity, $\omega$ is the angular frequency and $\epsilon_o$ is the permittivity of free space. The Debye parameters were fitted to the measured properties by using a weighted least squares method (W-LSM)[36] that was developed in MATLAB (R2017b, The MathWorks, Inc., Natick, MA, US) to fit a two-pole Debye model[25] for each tissue. The W-LSM assigns weights to the data points being fitted to emphasise certain points in the fitting process, depending on their relative importance[36]. As part of the fitting algorithm a weight factor $\gamma$ facilitates the control of the accuracy. The error was chosen to be dependent on the permittivity itself and is modelled as:

$$\dot{E}_i \approx \{\hat{\varepsilon}(\omega_i)\}^\gamma, \tag{4}$$

where $\hat{\varepsilon}(\omega_i)$ is the measured complex permittivity at the frequency $\omega_i$. The weight factor in this study was kept at 1 to assign equal weight to all permittivity values. The Newton iterative method to minimise the total weighted and the squared error is as follows[25]:

$$E^2 = \sum_{i=1}^{N_f} \left[ \frac{\{C_r(\omega_i) - M_r(\omega_i)\}^2}{\{E_r(\omega_i)\}^2} + \frac{\{C_i(\omega_i) - M_i(\omega_i)\}^2}{\{E_i(\omega_i)\}^2} \right] \tag{5}$$

where $N_f$ denotes the number of frequency points, $C_r(\omega_i)$ and $C_i(\omega_i)$ are the real and imaginary components of the calculated complex permittivity respectively, $M_r(\omega_i)$ and $M_i(\omega_i)$ are the real and imaginary components of the measured complex permittivity respectively and $E_r(\omega_i)$ and $E_i(\omega_i)$ are the real and imaginary components of the allowable error $\dot{E}_i$, respectively. Two-pole Debye models were fitted to the mean dielectric properties of all considered adrenal tissues (medulla, cortex, capsule, and tumour) over the frequency range (0.5–8.5 GHz). The mean square error (MSE) was calculated using the mean measured dielectric properties and the dielectric properties resulting from the two-pole Debye model. The MSE is calculated as follows:

$$MSE = \frac{1}{n} \times (\sum (\varepsilon_i - \hat{\varepsilon}_i)^2) \quad (6)$$

where $n$ denotes the number of data points, $\varepsilon_i$ is the $i_{th}$ actual value, $\hat{\varepsilon}_i$ is the $i_{th}$ predicted value and $\Sigma$ denotes the sum of the squared differences between predicted and actual values.

| Parameter | Uncertainty ($UC$) (%) | | Probability distribution | Divisor ($DIV$) | Standard uncertainty (%) $UC_{STD} = \frac{UC_{(Acc/REP)}}{DIV}$ |
|---|---|---|---|---|---|
| | Parameter | Value | | | |
| Relative permittivity | $UC_{ACC}$ | 0.88 | R | $\sqrt{3}$ | 0.50 |
| | $UC_{REP}$ | 0.16 | N | 1 | 0.16 |
| | $UC_{COM}$ | 0.52 | $\sqrt{\sum UC_{STD}^2}$ | – | – |
| | $UC_{EXT}$ (λ = 2) | 1.04 | $UC_{COM} \times \lambda$ | – | – |
| Conductivity | $UC_{ACC}$ | 4.35 | R | $\sqrt{3}$ | 2.51 |
| | $UC_{REP}$ | 1.24 | N | 1 | 1.24 |
| | $UC_{COM}$ | 2.79 | $\sqrt{\sum UC_{STD}^2}$ | – | – |
| | $UC_{EXP}$ (λ = 2) | 5.59 | $UC_{COM} \times \lambda$ | – | – |

**Table 2.** Uncertainty in accuracy and repeatability components for measured relative permittivity and conductivity of 0.1 M NaCl for a frequency range of 0.5–8.5 GHz. λ = multiplying factor for computing the expanded uncertainty, R = rectangular distribution, N = normal distribution.

## Results and discussion

This section presents the ex vivo dielectric properties of human adrenal glands. The dielectric properties of the medulla, cortex, and tumour are separately presented, to understand the variations in dielectric properties within each adrenal tissue. This individual analysis will provide an intra-tissue comparison of the dielectric properties of each tissue obtained from different adrenal diagnoses. Consequently, the measured dielectric properties of the cortex and tumours in the current study are compared with the existing literature. Finally, the two-pole Debye model parameters for the mean dielectric properties of the medulla, cortex, capsule, and tumour for different adrenal disorders are presented.

### Dielectric properties of medulla

The relative permittivity and conductivity of the medulla are shown in Fig. 3a,b respectively for patients diagnosed with Conn's syndrome, Cushing's syndrome, and Pheochromocytoma. The solid curves in Fig. 3 show the mean dielectric properties across different measurement locations of the medulla for each patient, while the error bars on each curve indicate the range of variation in dielectric properties across different measurement locations. It can be observed from Fig. 3a that the patients with Cushing's syndrome exhibit the highest relative permittivity of the medulla, as compared to those with Conn's syndrome and Pheochromocytoma.

The significant differences observed in the dielectric properties of the medulla between Conn's syndrome, Cushing's syndrome, and Pheochromocytoma can be attributed to the presence of lipids observed in the medulla of patients with Conn's syndrome. Contrary to this, no evidence of lipid composition was observed in the medulla of patients diagnosed with Cushing's syndrome and Pheochromocytoma. Additionally, changes in the tissue composition and structure of the medulla associated with these medical conditions are likely to cause variations in dielectric properties. The average percentage difference between the mean relative permittivity and conductivity of the medulla for each diagnosis is tabulated in Table 3.

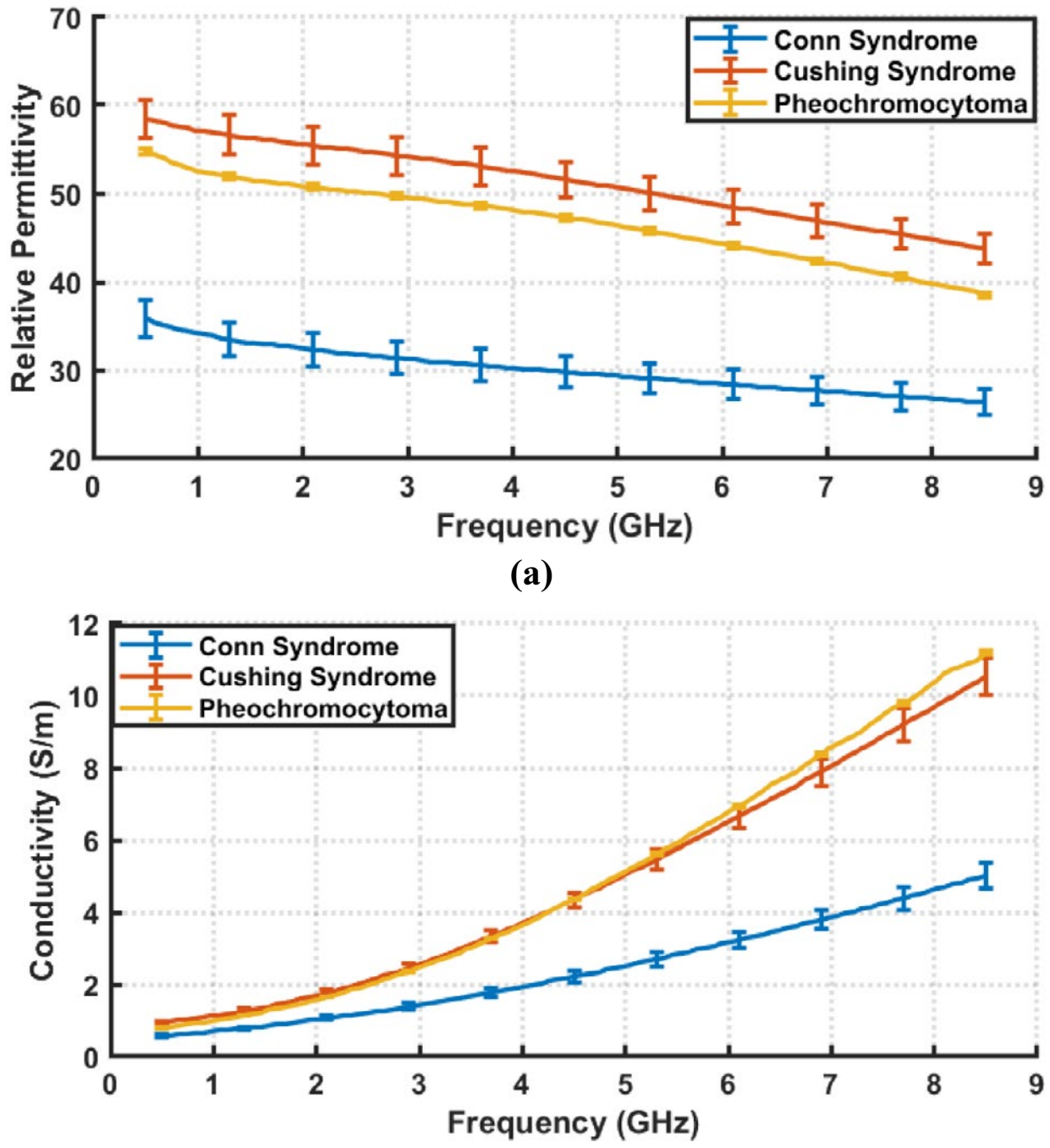

(a)

(b)

**Fig. 3.** Dielectric properties (**a**) Relative Permittivity (**b**) Conductivity for the medulla of ex vivo human adrenal glands.

| Diagnosis | Average percentage difference $\varepsilon_r$ (%) | Average percentage difference σ (%) |
|---|---|---|
| Cushing's syndrome versus pheochroersumocytoma | 9 | 5 |
| Cushing's syndrome versus Conn's syndrome | 52 | 61 |
| Conn's syndrome versus pheochromocytoma | 43 | 60 |

**Table 3.** Average percentage difference between the mean dielectric properties (relative permittivity ($\varepsilon_r$) and conductivity ($\sigma$ (S/m))) of the medulla for each diagnosis.

## Dielectric properties of cortex

The relative permittivity and conductivity of the cortex are shown in Fig. 4a, b for patients diagnosed with Conn's syndrome, Cushing's syndrome, and Pheochromocytoma. It can be observed from Fig. 4a that the patient with Pheochromocytoma exhibits the highest relative permittivity and conductivity of the cortex, as compared to those with Conn's syndrome and Cushing's syndrome. Further, a comparative analysis of the dielectric properties of the human adrenal cortex from the literature is presented in Fig. 4. Since the dielectric properties are highly species-dependent[37], the results are compared only with measurements of Bottiglieri et al.[18] who reported the ex vivo dielectric properties of the human adrenal cortex in the frequency range of 0.5–8 GHz. It can be observed from Fig. 4a that the relative permittivity of the human adrenal cortex reported by Bottiglieri et al.[18] agrees with the relative permittivity of the cortex of the Pheochromocytoma patient reported in this study, however, the conductivity values showed a significant difference. The results showed that the average percentage difference between the two studies was 5% for mean relative permittivity and 19% for conductivity.

The variation in the dielectric properties of the cortex between Conn's syndrome, Cushing's syndrome, and Pheochromocytoma could be attributed to their different underlying causes. Conn's syndrome results from an excess of aldosterone, which regulates salt and water balance, in the adrenal glands[38]. The overproduction of aldosterone can lead to changes in the composition of the water content and salts in the adrenal gland, including an increase in the number and size of cells in the zona glomerulosa of the adrenal cortex[39,40]. On the other hand, Cushing's syndrome results from an overproduction of cortisol, which has several functions in the body, by either the adrenal or pituitary gland[39]. The overproduction of cortisol can lead to changes in the structure and composition of the adrenal gland, including an increase in the size and number of cells in the zona fasciculate and zona reticularis of the adrenal cortex[40]. Pheochromocytoma is a rare condition caused by tumours in the adrenal gland or the sympathetic nervous system that produce excessive amounts of catecholamines, hormones that regulate blood pressure and heart rate[38,39]. Both Conn's syndrome and Cushing's syndrome primarily affect the adrenal cortex[38], whereas Pheochromocytoma primarily affects the adrenal medulla. Overall, the underlying changes in the tissue composition and structure in these medical conditions can lead to variations in the dielectric properties of the adrenal cortex. The average percentage difference between the mean relative permittivity and conductivity of the cortex for each diagnosis is tabulated in Table 4.

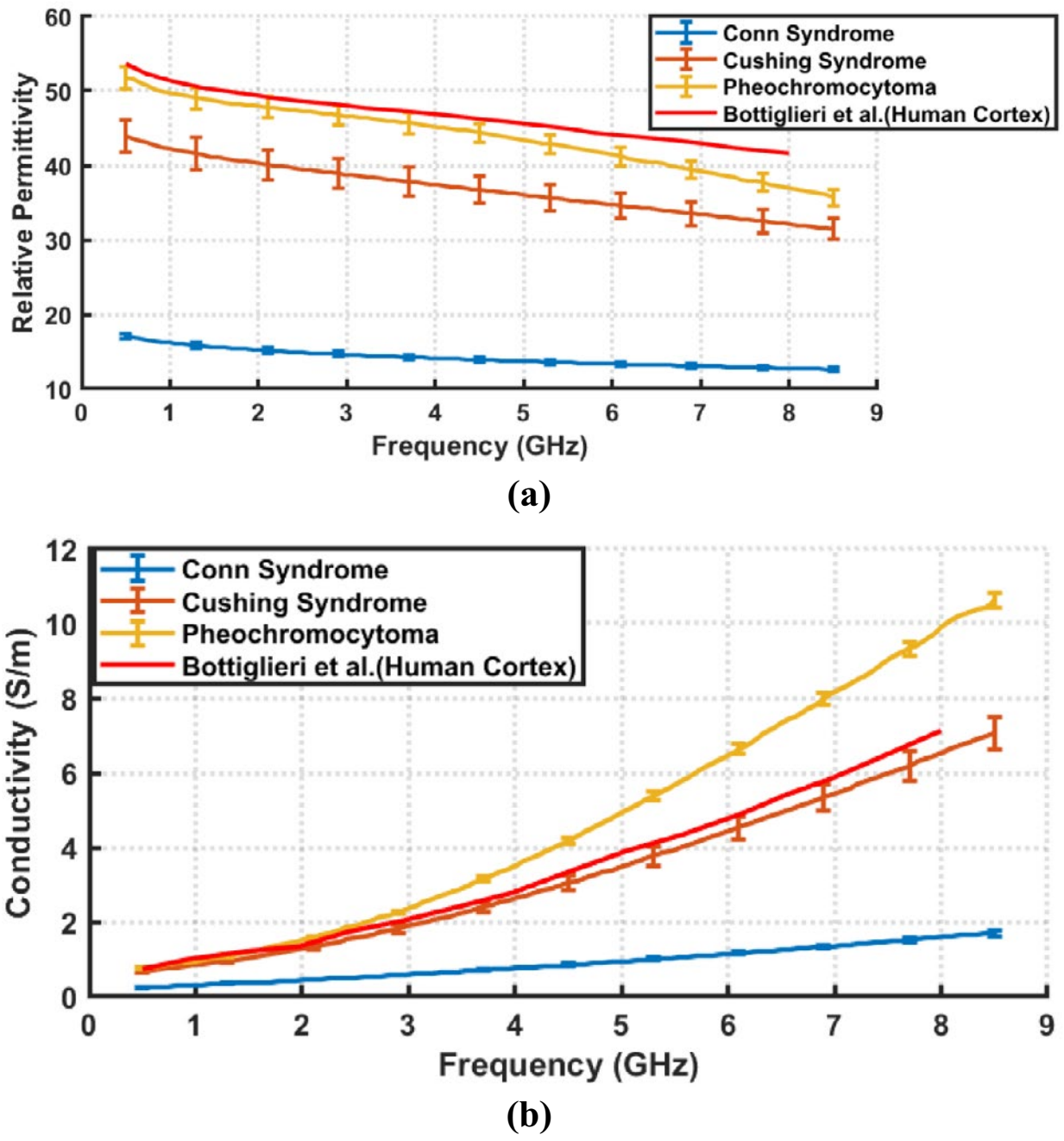

**Fig. 4.** Dielectric properties (**a**) Relative Permittivity (**b**) Conductivity for the cortex of ex vivo human adrenal glands.

## Dielectric properties of adrenal tumours

The relative permittivity and conductivity of the tumours are shown in Fig. 5a,b for patients diagnosed with Conn's syndrome, Cushing's syndrome, and Pheochromocytoma, respectively. It can be observed from Fig. 5a that the Pheochromocytoma tumour exhibits the highest relative permittivity and conductivity of the tumour, as compared to tumours of Conn's syndrome and Cushing's syndrome. Further, a comparative analysis of the dielectric properties of ex vivo human aldosterone-producing adenoma and Pheochromocytoma from the literature is presented in Fig. 5. Bottiglieri et al.[18] have reported the dielectric properties of human aldosterone-producing adenoma and Pheochromocytoma in the frequency range of 0.5–8 GHz. The dielectric properties of the human aldosterone-producing adenoma reported by Bottiglieri et al.[18] are consistent with the results of this study, with an average percentage difference of 10% for mean relative permittivity and 9% for conductivity. However, there is a notable difference in the dielectric properties of Pheochromocytoma reported by Bottiglieri et al.[18] compared to the results of this study, with an average percentage difference of 17% for mean relative permittivity and 29% for conductivity. The study by Bottiglieri et al.[18] did not specify whether the reported dielectric properties were for the Pheochromocytoma tumour or the cortex of Pheochromocytoma samples. In contrast, the current study reported the dielectric properties specifically for the Pheochromocytoma tumour. Consequently, the observed differences in the reported results between these studies could be attributed to variations in tissue type or inter-patient variation.

| Diagnosis | Average percentage difference $\varepsilon_r$ (%) | $\sigma$ (%) |
|---|---|---|
| Cushing's syndrome versus pheochromocytoma | 17 | 29 |
| Cushing's syndrome versus conn's syndrome | 89 | 110 |
| Conn's syndrome versus pheochromocytoma | 102 | 128 |

**Table 4.** Average percentage difference between the mean dielectric properties (relative permittivity ($\varepsilon_r$) and conductivity ($\sigma$ (S/m))) of the cortex for each diagnosis.

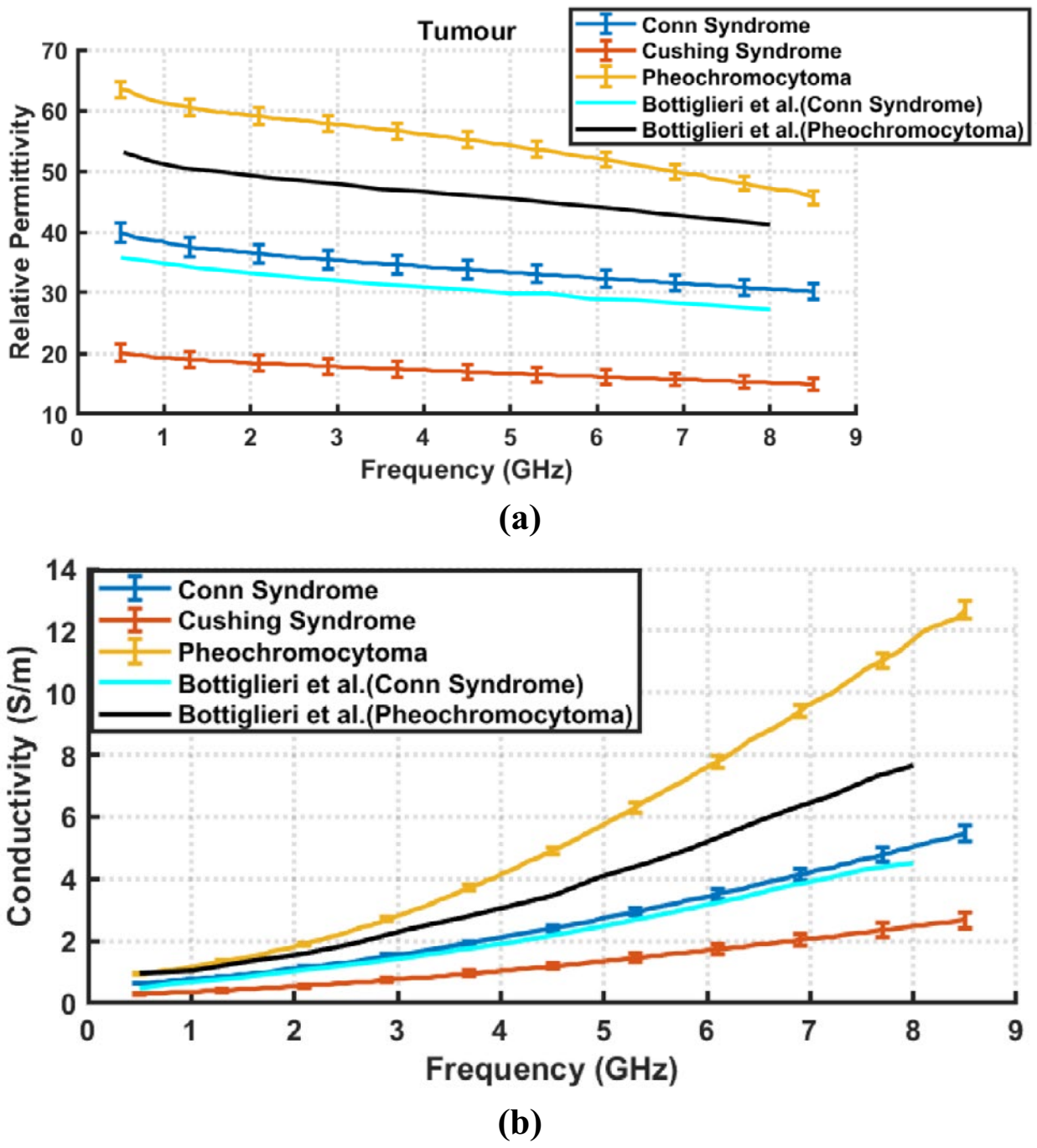

**Fig. 5.** Dielectric properties (**a**) Relative Permittivity (**b**) Conductivity for the tumours of ex vivo human adrenal glands.

The significant difference in the dielectric properties between Conn's syndrome and Pheochromocytoma tumours is likely due to the abundant lipid composition characteristic of aldosterone-producing adenomas[18]. One of the causes of Cushing's syndrome is due to adrenal lipid-rich adenoma (ALRA)[41]. ALRAs are characterised by their high lipid content[41], which likely contributes to the significantly lower dielectric properties of Cushing syndrome tumours compared to Pheochromocytoma and Conn's syndrome tumours. While the presence of lipids is a plausible explanation of low dielectric properties, it is important to further investigate tumour constituents to completely understand the primary factor influencing their dielectric properties. The average percentage difference between the mean relative permittivity and conductivity of the cortex for each diagnosis is tabulated in Table 5.

## Dielectric properties of capsule

The relative permittivity and conductivity of the capsule are shown in Fig. 6a,b for the patient diagnosed with Conn's syndrome. The patients diagnosed with Cushing's syndrome and Pheochromocytoma did not exhibit the presence of the capsule. The dielectric properties of the capsule were found to be higher when compared to the dielectric properties of the medulla in Conn's syndrome. The differences in physiological functions between the adrenal medulla and capsule can lead to differences in their cellular and molecular composition[42], which in turn can affect their dielectric properties. For example, the chromaffin cells of the adrenal medulla contain large amounts of catecholamines and other neurotransmitters[42], which can affect the dielectric properties of the tissue. In contrast, the adrenal capsule is primarily composed of collagen and other connective tissue components[43], which likely have different dielectric properties compared to the cells of the adrenal medulla. Therefore, the differences in dielectric properties between the adrenal capsule and medulla in Conn's syndrome are likely due to both the pathological changes occurring in the gland and the inherent differences in their physiological functions and cellular composition.

| Diagnosis | Average percentage difference | |
|---|---|---|
| | $\varepsilon_r$ (%) | $\sigma$ (%) |
| Cushing's syndrome versus pheochromocytoma | 105 | 120 |
| Cushing's syndrome versus Conn's syndrome | 66 | 69 |
| Conn's syndrome versus pheochromocytoma | 46 | 65 |

**Table 5.** Average percentage difference between the mean dielectric properties (relative permittivity ($\boldsymbol{\varepsilon_r}$) and conductivity ($\boldsymbol{\sigma}$ (S/m))) of the tumour for each diagnosis.

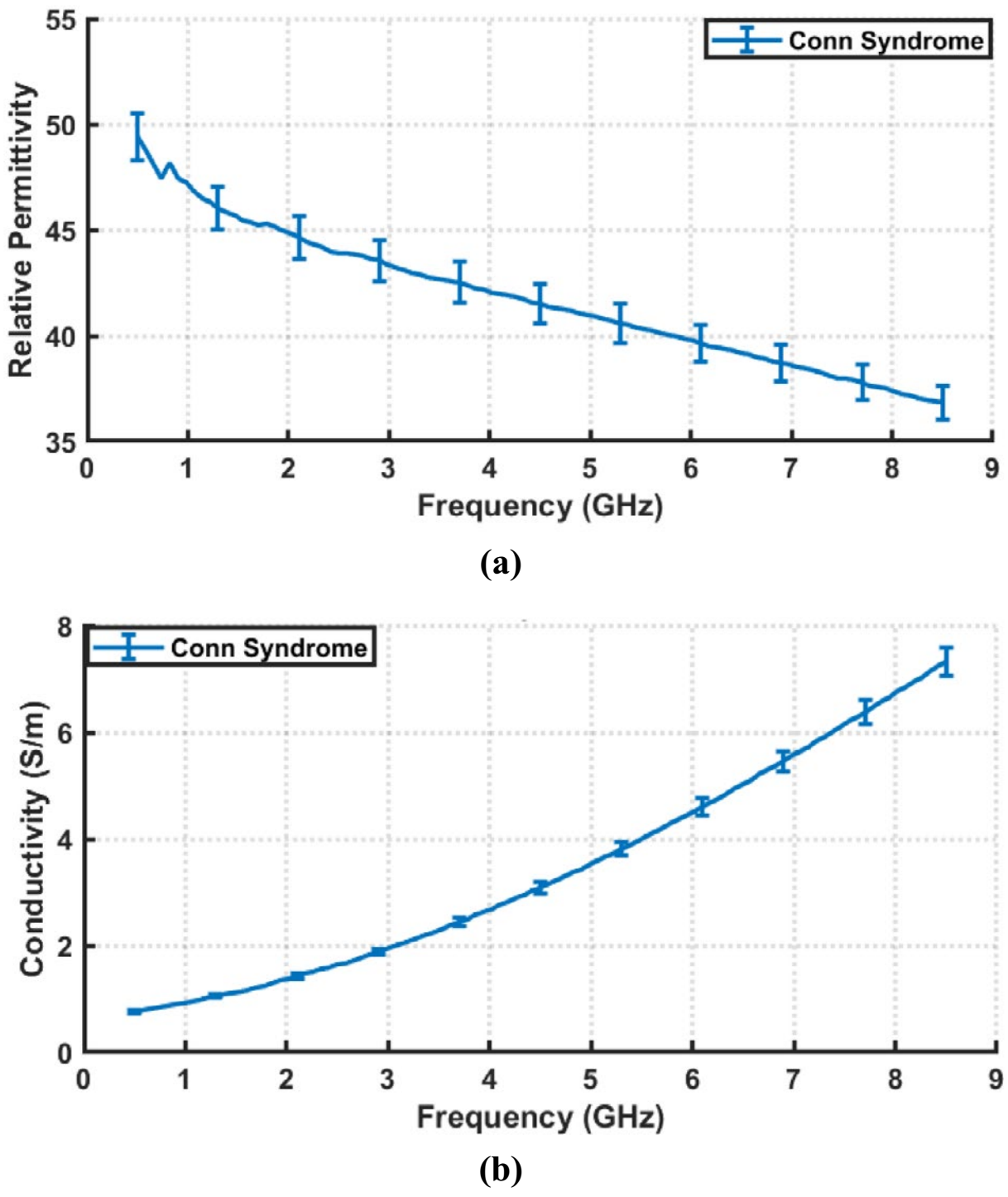

**Fig. 6.** Dielectric properties (**a**) Relative Permittivity (**b**) Conductivity for the capsule of ex vivo human adrenal gland.

## Fitting model on ex vivo human dielectric properties data

The Debye model is a mathematical model that describes the behaviour of dielectric properties of a material as a function of frequency. The model is commonly used to model biological tissues due to their complex dielectric behaviour, which is influenced by the presence of various molecular and cellular structures[25,44]. This study employed the two-pole Debye model to fit the mean dielectric properties of various components of the human adrenal gland, including the medulla, cortex, tumours, and capsule. Table 6 presents the parameters derived from fitting the two-pole Debye model to the mean dielectric properties of adrenal tissues under different diagnosed conditions. The findings indicate that the two-pole Debye model effectively captures the complex dielectric properties of adrenal tissues. These parameter insights will facilitate the development of EM diagnostic tools, including imaging techniques, for adrenal gland tissues through broadband EM simulations.

This study employed the MSE as a performance metric to assess the difference between the measured dielectric properties and those predicted by the two-pole Debye model across all frequency points examined in adrenal tissues. The tabulated results in Table 7 depict the MSE values. The MSE was computed for both relative permittivity and conductivity values across the investigated adrenal tissues. Table 7 shows that the MSE values vary according to tissue type and diagnosis. Furthermore, it is evident from Table 7 that the model demonstrated a strong alignment with the experimental data, as indicated by low MSE values for both relative permittivity and conductivity. The minimum MSE (0.00009) between the measured dielectric properties and the fitted model was observed for the conductivity of the adrenal cortex of Conn's syndrome, whereas the maximum MSE (0.1034) was observed for the relative permittivity of the Pheochromocytoma tumour.

The frequencies used in MWA are important because they determine the penetration depth, heating pattern, and other characteristics of the EM waves, and thus impact the efficacy and safety of the ablation procedure[45]. The most commonly used frequencies in MWA are 915 MHz, 2.45 GHz, and 5.80 GHz[18,46]. Understanding the dielectric properties of adrenal tissues at these frequencies will help in determining how efficiently the adrenal tissues can absorb and convert EM energy into heat, which is essential for MWA for adrenalectomy. Therefore, understanding the dielectric properties of tissues at these frequencies is crucial for selecting the appropriate frequency for MWA and optimizing the effectiveness and safety of the procedure. To this end, the mean dielectric properties of all investigated adrenal tissues of each diagnosis are tabulated in Table 8.

| Diagnosis | Tissue | $\varepsilon_\infty$ | $\sigma_s\,(S/m)$ | $\Delta\varepsilon_1$ | $\tau_1$ (ps) | $\Delta\varepsilon_2$ | $\tau_2$ (ps) |
|---|---|---|---|---|---|---|---|
| Conn's syndrome | Medulla | 11.93 | 0.50 | 19.75 | 11.60 | 4.30 | 142.75 |
| | Cortex | 8.69 | 0.21 | 2.65 | 134.24 | 5.85 | 13.14 |
| | Tumour | 14.12 | 0.56 | 21.76 | 11.47 | 4.14 | 144.71 |
| | Capsule | 13.30 | 0.65 | 30.84 | 10.72 | 5.79 | 166.11 |
| Cushing's syndrome | Medulla | 13.20 | 0.07 | 435.08 | 4465.80 | 43.52 | 12.54 |
| | Cortex | 11.43 | 0.60 | 28.28 | 12.20 | 4.62 | 146.01 |
| | Tumour | 7.75 | 0.25 | 2.16 | 141.80 | 10.30 | 12.45 |
| Pheochromocytoma | Medulla | 1.00 | 0.56 | 50.56 | 10.93 | 17.62 | 609.53 |
| | Cortex | 1.00 | 0.62 | 47.51 | 11.19 | 6.52 | 352.53 |
| | Tumour | 4.88 | 0.03 | 55.61 | 11.35 | 266.74 | 2661.95 |

**Table 6.** Two-pole Debye model parameters of all measured tissues for all diagnoses.

| Diagnosis | Tissue | Mean-squared error (MSE) | |
|---|---|---|---|
| | | $\varepsilon_r$ | $\sigma$ |
| Conn's syndrome | Medulla | 0.0327 | 0.0003 |
| | Cortex | 0.0144 | 0.00009 |
| | Tumour | 0.0340 | 0.0003 |
| | Capsule | 0.0535 | 0.0005 |
| Cushing's syndrome | Medulla | 0.0694 | 0.0010 |
| | Cortex | 0.0323 | 0.0004 |
| | Tumour | 0.0087 | 0.0001 |
| Pheochromocytoma | Medulla | 0.0148 | 0.0030 |
| | Cortex | 0.0260 | 0.0026 |
| | Tumour | 0.1034 | 0.0057 |

**Table 7.** Mean-squared error value of the fit error. The error values for the fit of each tissue type are calculated as the mean value across all frequency points.

| Diagnosis | Medulla | | Cortex | | Tumour | | Capsule | |
|---|---|---|---|---|---|---|---|---|
| | $\varepsilon_r$ | $\sigma$ (S/m) | $\varepsilon_r$ | $\sigma$ (S/m) | $\varepsilon_r$ | $\sigma$ (S/m) | $\varepsilon_r$ | $\sigma$ (S/m) |
| 915 MHz | | | | | | | | |
| Conn's syndrome | 34.16 | 0.17 | 16.16 | 0.08 | 38.23 | 0.17 | 47.05 | 0.24 |
| Cushing's Syndrome | 57.15 | 1.02 | 42.28 | 0.21 | 19.29 | 0.09 | – | – |
| Pheochromocytoma | 52.68 | 0.39 | 49.59 | 0.28 | 61.38 | 1.06 | – | – |
| 2.45 GHz | | | | | | | | |
| Conn's syndrome | 31.80 | 0.69 | 14.81 | 0.30 | 35.91 | 0.72 | 44.09 | 0.95 |
| Cushing's syndrome | 55.24 | 1.96 | 39.50 | 0.94 | 18.05 | 0.37 | – | – |
| Pheochromocytoma | 50.36 | 1.39 | 47.35 | 1.25 | 59.00 | 2.17 | – | – |
| 5.80 GHz | | | | | | | | |
| Conn's syndrome | 28.83 | 2.59 | 13.55 | 0.92 | 32.78 | 2.80 | 40.21 | 3.74 |
| Cushing's syndrome | 49.21 | 6.17 | 35.20 | 3.72 | 16.36 | 1.40 | – | – |
| Pheochromocytoma | 44.67 | 5.92 | 41.77 | 5.58 | 52.39 | 7.22 | – | – |

**Table 8.** Mean dielectric properties (relative permittivity ($\varepsilon_r$) and conductivity ($\sigma$ (S/m))) of all measured tissues for all diagnoses at 915 MHz, 2.45 GHz, and 5.80 GHz.

## Conclusions

This study has reported the ex vivo dielectric properties of human adrenal glands obtained from four different patients who underwent unilateral adrenalectomy at the University Hospital Galway, Ireland. Two of the patients included in this study were diagnosed with Cushing's syndrome, while the remaining two patients were diagnosed with Conn's syndrome and Pheochromocytoma. The dielectric properties were characterised for a frequency range of 0.5–8.5 GHz by using an OECL probe. To the best of the knowledge of the authors, this study has documented first time the dielectric properties of the human adrenal medulla, capsule and different tumour types associated with adrenal glands from patients diagnosed with Conn's syndrome, Cushing's syndrome, and Pheochromocytoma. A two-pole Debye model was fitted on the dielectric properties of the investigated adrenal tissues from all three different diagnosed adrenal disorders. The dielectric properties of each of these tissues from all four patients were compared at three different frequency points including 915 MHz, 2.45 GHz, and 5.8 GHz. These frequencies are commonly used for designing an interstitial antenna for the application of MWA.

The results of this study indicate that there exists a significant difference in the dielectric properties of tumours from patients with Conn's syndrome, Cushing's syndrome, and Pheochromocytoma, with Cushing's syndrome tumour exhibiting the lowest dielectric properties due to its lipid-rich composition compared to Conn's syndrome and Pheochromocytoma tumours. The dielectric properties of the adrenal medulla and cortex also show a notable difference among patients diagnosed with these three syndromes, attributed to the distinct nature of diseases as Conn's and Cushing's syndrome affect the adrenal cortex while Pheochromocytoma affects the adrenal medulla. Moreover, patients with Cushing's syndrome typically experience more severe hypokalemia and hypernatremia than those with Conn's syndrome. Consequently, the variation in dielectric properties between patients with these two syndromes is likely due to the electrolyte imbalances caused by these conditions. Pheochromocytoma, on the other hand, arises from tumours in the adrenal medulla that secrete excess amounts of catecholamines, which can affect the dielectric properties of the adrenal medulla.

The dielectric properties of human adrenal glands presented in this paper can be utilised to create precise anatomical models to assess new medical imaging and therapeutic devices for addressing adrenal gland-induced hypertension. Moreover, these properties will be useful in developing and optimising MWA therapy for adrenal tumours. While this paper reported ex vivo dielectric properties of human adrenal glands, it would be more insightful to perform the in vivo measurements to better understand how the dielectric properties of the glands change under physiological conditions. Additionally, further comprehensive studies are necessary to confirm these preliminary findings and to better understand the influence of patient-specific variables on the dielectric properties of human adrenal glands. Future research with an expanded patient cohort will allow for a more robust analysis, potentially leading to more definitive conclusions regarding the impact of age, sex, and other factors on tissue dielectric properties. The accurate characterisation of these properties will help in designing an optimal treatment protocol that delivers the appropriate amount of energy to the tumour while minimising damage to surrounding healthy tissue.

## Data availability

All data generated or analysed during this study are included in this published article [and its supplementary information files].

## Acknowledgements
This publication has emanated from research conducted with the financial support of Science Foundation Ireland (20/US/3676), the European Regional Development Fund (ERDF) under grant number 13/RC/2073_2, the National Institutes of Health (NIH) under grant number R01EB028848, and Future*Care* Seed Fund 2023 by College of Medicine, Nursing and Health Sciences, University of Galway, Ireland.

## Author contributions
B. A., A. E., G. C., and M. O. conceived the experiment(s), B. A., G. C., N. I., A. M., A. L. conducted the experiment(s), B. A., A. E., E. D., F. K., analysed the results, B. A. wrote the original draft, A. L., P. P., M. C., and M. O. arranged the resources to conduct the experiment(s). All authors reviewed the manuscript.

## Competing interests
The authors declare no competing interests.

## Additional information
**Supplementary Information** The online version contains supplementary material available at https://doi.org/10.1038/s41598-024-70476-6.

**Correspondence** and requests for materials should be addressed to B.A.